\documentclass[11pt,a4paper]{article}

\usepackage{amsmath,amssymb,amsthm}
\usepackage{hyperref}
\usepackage{geometry}
\usepackage{cite}

\title{\textbf{Curvature of Physical Space and Geometric Projection: \\ A Possible Geometric Understanding of Quantum Mechanics and Relativity}}
\author{Xiang Qi}
\date{}

\begin{document}
\maketitle

% ========== Abstract ==========
\begin{abstract}
Starting from a formal comparison between the Schr\"{o}dinger equation and the Jacobi field equation, this paper proposes the hypothesis that, within the framework of non-relativistic quantum mechanics, the sectional curvature of microscopic physical space can be expressed as \(K(x) = 2m(E-V(x))/\hbar^2\), and attempts to offer a possible unified geometric understanding of quantum mechanics and relativity. The paper is divided into three parts. The first part identifies the Schr\"{o}dinger equation, in its mathematical form, as a Jacobi field equation. The wave function is made to correspond to a Jacobi field, giving it the geometric meaning of the deviation of geodesics in physical space. Phenomena such as quantum tunneling, energy quantization, and the path integral are then interpreted, from this perspective, as geometric manifestations of different spatial curvatures. The second part proposes a geometric projection scheme based on semi-geodesic coordinates. The transverse metric factor of a curved space, which satisfies a Jacobi field equation, is taken as a unified geometric framework. It is demonstrated that the metric correction terms in classical mechanics, special relativity, general relativity, and quantum mechanics can all be derived from the same equation, with the curvature \(K\) arising from different physical sources. The third part extends this framework to statistical mechanics.
\end{abstract}

% ========== Main Text ==========
\section{Introduction}

Quantum mechanics is one of the most successful theories of twentieth-century physics. Its central equation, the Schr\"{o}dinger equation \cite{Schrodinger1926}, describes the behavior of microscopic particles with extraordinary precision. Yet, within the standard framework, while phenomena such as tunneling, path integrals, the uncertainty principle, and energy quantization are derived with mathematical rigor, the question of whether they can be understood from a more unified geometric perspective remains an open and worthwhile inquiry. Concurrently, general relativity attributes gravitational phenomena to the intrinsic curvature of spacetime \cite{Einstein1915}, demonstrating that fundamental physical interactions can be described in a geometric language. Whether a connection exists between the geometric foundations of quantum mechanics and relativity is also a topic ripe for exploration.

This paper attempts to offer a possible unified geometric understanding of quantum mechanics and relativity. The operational approach adopted herein draws upon a methodological tradition that has repeatedly proven fruitful in the history of physics: establishing an equivalence between a physical quantity characterizing a system's state and a geometric quantity describing the structure of space, and then allowing geometric theorems to complete the remaining work.

This tradition can be traced back to Hamilton and Jacobi. In 1834--1835, Hamilton introduced the concept of generalized momentum \cite{Hamilton1834,Hamilton1835}, transforming mechanical equations into a first-order canonical form and endowing classical mechanics with a geometric meaning in phase space. In his series of works from 1836 to 1837, Jacobi \cite{Jacobi1836,Jacobi1837} further proposed Jacobi's principle, showing that for systems with conserved energy, a particle's trajectory in space is equivalent to a geodesic under an effective Riemannian metric, \(\tilde{g}_{ij} = 2m(E-V)a_{ij}\), thereby linking the physical quantity of kinetic energy with the geometric quantity of the metric.

Einstein elevated this methodology to a new level. In the final formulation of general relativity in 1915 \cite{Einstein1915}, he proposed the field equation \(G_{\mu\nu} = 8\pi G T_{\mu\nu}\), which directly relates the physical distribution of matter to the geometric curvature of spacetime, providing a fundamentally new geometric framework for understanding gravity. Minkowski's formulation of special relativity in terms of four-dimensional spacetime geometry \cite{Minkowski1909} also embodies this same methodological characteristic.

This methodology finds a certain continuation in modern gauge field theory. The Yang-Mills field equation, proposed by Yang and Mills in 1954 \cite{YangMills1954}, can be written in modern fiber bundle language as \(D_\mu F^{\mu\nu} = J^\nu\). The left-hand side, the covariant divergence of the fiber bundle curvature, describes the bending of the gauge space, while the right-hand side, the matter field current, describes the distribution of the physical source. This structure once again equates a geometric quantity with a physical one.

A profound resonance of this methodology also emerged in the field of statistics. Rao first noted in 1945 that the Fisher information matrix possesses the mathematical properties of a Riemannian metric \cite{Rao1945}. This was systematically developed by Amari beginning in the 1980s \cite{Amari1985}, who established a correspondence between the statistical distinguishability of probability distribution families and a Riemannian metric, providing a differential geometric framework for statistical inference.

From Hamilton to Amari, these works, spanning two centuries, share a deep operational commonality: they all directly equate a quantity that profoundly characterizes a system's state with a geometric quantity. The present work follows a similar methodological principle, namely an attempt to recognize geometrically the mathematical structures already present in the standard equations. Following this tradition, this paper proceeds from the Schr\"{o}dinger equation to propose a curvature hypothesis, \(K(x)= 2m(E-V(x))/\hbar^2\), which establishes a correspondence between the physical quantity of energy difference and the geometric quantity of sectional curvature. Under this hypothesis, many phenomena of quantum mechanics, as well as statistical mechanics, may acquire a unified geometric understanding. This paper further introduces the geometric conclusion of a transverse metric factor arising from the projection of a curved space in semi-geodesic coordinates, thereby extending this geometric framework to relativity.

It should be stressed that the curvature hypothesis originates from a rearrangement of the time-independent Schr\"{o}dinger equation and therefore constitutes a geometric reinterpretation of non-relativistic quantum mechanics. Consequently, the geometric interpretation of quantum-mechanical phenomena in this paper is confined to the non-relativistic domain. A preliminary exploration of the corresponding generalization to relativistic quantum mechanics has been carried out in a subsequent work~\cite{Qi2026_YangMills}.

% ==================== Part I: Geometric Interpretation of Quantum Mechanics ====================

\section{Derivation of the Physical Space Curvature}

Before proceeding to the concrete geometric interpretations, it is necessary to clarify the assumptions made in this paper regarding the analytical method, dimensionality, and nature of the work.

\textbf{First, on the use of the time-independent Schr\"{o}dinger equation.} The time-independent Schr\"{o}dinger equation is the standard tool for treating quantum systems where the potential energy is not explicitly time-dependent. This work focuses on stable properties of the system, such as energy levels, tunneling probabilities, and uncertainty relations, all of which are described by the time-independent equation. As time evolution can be obtained by linear superposition of stationary states, the analysis of the stationary case does not sacrifice generality.

\textbf{Second, on treating the potential function as locally constant.} When analyzing specific quantum phenomena, this paper approximates the potential \(V(x)\) as constant within a local region, thereby treating the curvature \(K(x)\) as constant and directly applying the standard solutions for constant-curvature spaces. This approach is consistent with the pedagogical practice of using piecewise constant potentials (e.g., finite square wells and barriers) in quantum mechanics, and its systematic development is known as the WKB approximation \cite{Landau1981,Griffiths2018}. In the geometric framework of this paper, this approximation corresponds to treating space locally as a constant-curvature space: solutions oscillate locally in regions of positive curvature, decay exponentially locally in regions of negative curvature, and are locally linear at points of zero curvature.

\textbf{Third, on the dimensionality setting of this paper.} The geometric model is set on a Riemannian manifold of at least two dimensions. The Jacobi field equation describes the transverse deviation of geodesics in the normal space. It is not required that the entire normal space be one-dimensional; rather, it is assumed that the Jacobi field is confined to a specific direction within the normal space. By taking a parallel-transported unit normal vector field along this direction, the Jacobi field equation reduces to the scalar form \(f'' + K f = 0\), where \(x\) is the arc length parameter along the geodesic, \(f(x)\) is the amplitude of the normal deviation, and \(K(x)\) is the sectional curvature of the plane spanned by the tangent vector and the chosen normal vector. All core derivations in this paper are based on this reduced scalar equation.

Under this assumption, the physical space must be at least two-dimensional, for only then does the sectional curvature \(K\) have a non-trivial geometric meaning. The spatial coordinate \(x\) in the time-independent Schr\"{o}dinger equation is identified here with the arc length parameter of a geodesic, the wave function \(\psi(x)\) is identified with the amplitude of the Jacobi field along that specific normal direction, and the potential \(V(x)\) varies along the geodesic. This identification is mathematically rigorous and does not conflict with the standard ``one-dimensional problem'' in quantum mechanics---the standard setting confines a particle to a line, while the present setting has a particle moving along a geodesic; the mathematical descriptions of the two are formally isomorphic.

The choice to focus on a specific normal direction is motivated primarily by the following considerations. First, constant-curvature spaces under this setting possess three standard elementary solutions (trigonometric, linear, and exponential functions), which can be directly compared with the correction terms in various physical theories, thereby making the geometric structure clearest. Second, this simplification is sufficient to demonstrate the core ideas of this paper: the correspondence between the wave function and the Jacobi field, the decisive role of the curvature's sign on the qualitative behavior of solutions, and the formal correspondence between the projection correction factor and standard physical correction terms. The Jacobi field equation is defined on Riemannian manifolds of any dimension and for normal spaces of any dimension \cite{doCarmo1992}. Generalization of this geometric framework to higher dimensions is mathematically direct: in higher dimensions, each independent direction in the normal space still satisfies a scalar equation of the same form. The concrete calculations and establishment of physical correspondences for such a generalization remain a task for future work.

\textbf{Fourth, on the nature of this work.} This paper aims to provide a possible geometric framework for understanding quantum mechanics and relativity, not to compete with standard theories in numerical precision. Standard quantum mechanics and relativity have been rigorously validated by numerous experiments within their respective domains of applicability. The geometric reinterpretation offered here is mathematically equivalent to these standard theories: the rearranged Schr\"{o}dinger equation and the Jacobi field equation are mathematically the same equation, and assigning a different physical interpretation to the solution of the same equation does not change its empirical predictions. Hence, the present geometric reinterpretation automatically inherits all of its experimental support.

\subsection{Formal Comparison of the Schr\"{o}dinger Equation and the Jacobi Field Equation}

This paper takes the one-dimensional time-independent Schr\"{o}dinger equation as its starting point \cite{Schrodinger1926}:
\begin{equation}
-\frac{\hbar^2}{2m} \frac{d^2\psi}{dx^2} + V(x)\psi = E\psi
\label{eq:schrodinger}
\end{equation}
where \(m\) is the mass of the particle, \(V(x)\) is the potential energy, \(E\) is the energy eigenvalue, and \(\psi(x)\) is the wave function.

Rearranging Equation (\ref{eq:schrodinger}) yields:
\begin{equation}
\frac{d^2\psi}{dx^2} + \frac{2m(E-V(x))}{\hbar^2} \psi = 0
\label{eq:schrodinger_rearranged}
\end{equation}

Now, turning to Riemannian geometry. On any Riemannian manifold, a Jacobi field \(J(s)\) along a geodesic \(\gamma(s)\) describes the deviation behavior of neighboring geodesics \cite{doCarmo1992}. Suppose there exists a parallel-transported unit normal vector field \(E(s)\) along \(\gamma(s)\), and let the component of the Jacobi field along this direction be \(J(s) = f(s)E(s)\). The Jacobi field equation then reduces to the following scalar equation for this normal component:
\begin{equation}
\frac{d^2 f}{ds^2} + K_{\dot{\gamma}, E}(s) \cdot f(s) = 0,
\label{eq:jacobi}
\end{equation}
where \(K_{\dot{\gamma}, E}(s) = \langle R(E, \dot{\gamma})\dot{\gamma}, E\rangle\) is the sectional curvature of the plane spanned by the tangent vector \(\dot{\gamma}\) and the normal vector \(E\), \(R\) is the Riemann curvature tensor, and \(s\) is the arc length parameter. For brevity, \(K_{\dot{\gamma}, E}(s)\) will be denoted simply as \(K(s)\) hereafter.

When \(K(s)\) is not constant, the solutions to Equation (\ref{eq:jacobi}) generally cannot be expressed as elementary functions, but their qualitative behavior is strictly controlled by the sign of the curvature via the \textbf{Sturm Comparison Theorem} \cite{Coddington1955}; for applications of this theorem in Riemannian geometry, see \cite{Cheeger1975}. Let \(f(s)\) be a non-zero solution to the equation; the following conclusions hold:
\begin{itemize}
\item If there exists a constant \(c > 0\) such that \(K(s) \ge c\) holds everywhere along \(\gamma\), then \(f(s)\) has at least one zero on any arc segment with a length greater than \(\pi/\sqrt{c}\). The solution exhibits oscillatory behavior, and the distance between adjacent zeros does not exceed \(\pi/\sqrt{c}\). In this case, conjugate points must exist along the geodesic.
\item If \(K(s) \le 0\) holds everywhere along \(\gamma\), then \(f(s)\) has at most one zero, and no conjugate points exist along the geodesic. The solution exhibits exponential divergence or decay, and no periodic oscillation occurs.
\end{itemize}

Furthermore, when the sectional curvature \(K\) is \textbf{constant}, Equation (\ref{eq:jacobi}) becomes a linear ODE with constant coefficients, and its exact general solution can be directly obtained:
\begin{itemize}
\item \textbf{Positive curvature \(K > 0\)}: The general solution is \(f(s) = A\cos(\sqrt{K}\,s) + B\sin(\sqrt{K}\,s)\). This represents simple harmonic oscillation with frequency \(\sqrt{K}\). The distance between adjacent zeros is exactly \(\pi/\sqrt{K}\). Geometrically, this means that neighboring geodesics emanating from a point must converge at a conjugate point after an arc length of \(\pi/\sqrt{K}\).
\item \textbf{Zero curvature \(K = 0\)}: The general solution is \(f(s) = As + B\). Neighboring geodesics separate (or approach) at a constant rate. This corresponds to the flat geometry of Euclidean space.
\item \textbf{Negative curvature \(K < 0\)}: Let \(\kappa = \sqrt{-K} > 0\). The general solution is \(f(s) = A\cosh(\kappa s) + B\sinh(\kappa s)\). The solution diverges exponentially. For \(s \gg 1/\kappa\), the behavior of \(f(s)\) is dominated by \(\frac{1}{2}(A+B)e^{\kappa s}\). Neighboring geodesics diverge rapidly at an exponential rate. A non-zero solution has at most one zero, and no conjugate points exist along the entire geodesic.
\end{itemize}

\subsection{The Curvature Hypothesis}

Comparing Equation (\ref{eq:schrodinger_rearranged}) with Equation (\ref{eq:jacobi}), the two are formally identical:
\begin{itemize}
    \item The wave function \(\psi(x)\) corresponds to the Jacobi field \(f(s)\);
    \item The spatial coordinate \(x\) corresponds to the arc length parameter \(s\);
    \item The coefficient term \(\frac{2m(E-V(x))}{\hbar^2}\) corresponds to the sectional curvature \(K(s)\) of the plane spanned by the wave function's direction and the spatial coordinate direction.
\end{itemize}

Based on this strict formal correspondence, this paper proposes the following core hypothesis: within the context of non-relativistic quantum mechanics, the coefficient term \(\frac{2m(E-V(x))}{\hbar^2}\) is identified as the sectional curvature of microscopic physical space, denoted by \(K(x)\). This hypothesis is the logical starting point for all derivations in this paper:
\begin{equation}
\boxed{K(x) = \frac{2m(E-V(x))}{\hbar^2}}
\label{eq:K_hypothesis}
\end{equation}

Equation (\ref{eq:schrodinger_rearranged}) can be rewritten as:
\begin{equation}
\frac{d^2\psi}{dx^2} + K(x) \psi = 0
\label{eq:schrodinger_jacobi}
\end{equation}
This equation is mathematically strictly equivalent to the Jacobi field equation (\ref{eq:jacobi}).

The core of hypothesis (\ref{eq:K_hypothesis}) lies not only in the formal mathematical correspondence but also in the physical claim it makes. The left-hand side is the sectional curvature \(K(x)\), describing the degree of bending in physical space. The right-hand side is a combination of the mass \(m\), the energy difference \(E-V(x)\), and the reduced Planck constant \(\hbar\), describing the dynamical state of a microscopic particle. This hypothesis proposes that the physical quantities on the right-hand side may determine the curvature of microscopic physical space. The curvature is jointly determined by these physical quantities, where the sign of the energy difference dictates the sign of the curvature: if the energy difference is positive (\(E>V(x)\)), the space has positive curvature; if it is negative (\(E<V(x)\)), the space has negative curvature; and if it is zero, the space is flat.

\textbf{Definition of the curvature radius.}
From the curvature hypothesis, a characteristic scale with the dimension of length can be naturally defined, namely the curvature radius:
\begin{equation}
R_c(x) \equiv \frac{1}{\sqrt{|K(x)|}} = \frac{\hbar}{\sqrt{2m|E-V(x)|}}
\label{eq:curvature_radius}
\end{equation}
In differential geometry, the curvature radius is a fundamental geometric quantity for describing the degree of bending of a surface. For a two-dimensional surface, the curvature radius is defined as the reciprocal of the absolute value of the Gaussian curvature \cite{doCarmo1992}; in higher-dimensional Riemannian manifolds, it is defined through the sectional curvature \cite{Petersen2016}. In the present context, the curvature radius carries a clear physical meaning: in a positive-curvature region (\(E > V(x)\)), \(R_c\) describes the characteristic scale of periodic geodesic convergence, i.e., the arc length between adjacent conjugate points is \(\pi R_c\); in a negative-curvature region (\(E < V(x)\)), \(R_c = 1/\kappa\) describes the characteristic length of exponential geodesic divergence or exponential wave-function decay, i.e., the tunneling depth; in a zero-curvature region (\(E = V(x)\)), the curvature radius diverges and the space is flat.

\subsection{Three Curvatures and Quantum Behavior}

In standard quantum mechanics, problems involving a specific potential \(V(x)\) are often approached by partitioning space into regions according to the relative magnitude of the energy \(E\) and the potential \(V(x)\). Specifically: \(E > V(x)\) defines the classically allowed region, where solutions to the Schr\"{o}dinger equation are oscillatory; \(E < V(x)\) defines the classically forbidden region, where solutions are exponentially decaying; and \(E = V(x)\) defines the classical turning point, where solutions transition between these two behaviors.

The contribution of this paper is: building upon this existing classification, and via the curvature hypothesis \(K(x) = 2m(E-V(x))/\hbar^2\), to identify these three physical situations with three distinct geometric states. Because the sign of the curvature is entirely determined by the energy difference \(E-V(x)\), the three types of regions demarcated in standard quantum mechanics according to the relation between \(E\) and \(V(x)\) correspond precisely to three spaces of distinct curvature in the geometric framework of this paper.

\textbf{Positive-curvature space (\(K(x) > 0\), i.e., \(E > V(x)\)).} This is the classically allowed region. In a positive-curvature space, solutions to the Jacobi field equation are trigonometric functions; geometrically, geodesics converge periodically. Physically, this corresponds to the oscillatory behavior of the wave function. When the system is subject to boundary conditions, the trigonometric solution must vanish at the boundaries, forcing the curvature to take only discrete values, which naturally leads to energy quantization.

\textbf{Zero-curvature space (\(K(x) = 0\), i.e., \(E = V(x)\)).} This is the classical turning point. In a zero-curvature space, the Jacobi field equation reduces to \(f'' = 0\), with a linear function as its solution. Geometrically, geodesics are uniformly distributed, neither converging nor diverging. Physically, this corresponds to the region where the WKB approximation fails and a precise connection using Airy functions is required. In the geometric framework of this paper, zero curvature is the critical state between positive and negative curvature.

\textbf{Negative-curvature space (\(K(x) < 0\), i.e., \(E < V(x)\)).} This is the classically forbidden region. In a negative-curvature space, solutions to the Jacobi field equation are exponential functions; geometrically, geodesics diverge from each other at an exponential rate. Physically, this corresponds to the exponentially decaying behavior of the wave function. From a geometric standpoint, the exponential divergence of neighboring geodesics in a negative-curvature space implies that geodesics emanating from nearby initial conditions become extremely dispersed. The particle has no single, uniquely determined classical path and contributions from multiple paths must be considered simultaneously, which corresponds to the characteristic behaviors of tunneling and the path integral in quantum mechanics.

Through the correspondences above, this paper provides a possible geometric reinterpretation of the existing classification of regions in quantum mechanics by the relation between \(E\) and \(V(x)\): the classically allowed region corresponds to a positive-curvature space; the classically forbidden region corresponds to a negative-curvature space; and the classical turning point corresponds to a zero-curvature space. This correspondence forms the geometric foundation upon which all subsequent discussions in this paper are built.

% ==================== Section 3: Geometric Interpretation of Standard Quantum Mechanical Formulas ====================

\section{Geometric Interpretation of Standard Quantum Mechanical Formulas}

The following physical conclusions are all standard results of quantum mechanics \cite{Landau1981,Sakurai2017}. This paper merely provides their reinterpretation within the geometrically equivalent framework based on the curvature hypothesis.

\subsection{Geometric Interpretation of the Tunneling Formula}

\textbf{Physical conditions and sign of curvature.} Quantum tunneling occurs when the particle's energy is less than the barrier height, i.e., \(E < V_0\). In this region, the curvature hypothesis \(K = 2m(E-V_0)/\hbar^2\) directly yields \(K < 0\), meaning the region is a negative-curvature space.

\textbf{Standard physics interpretation.} In quantum mechanics, the wave function inside the potential barrier takes the form of an exponential decay, \(\psi(x) \propto e^{-\kappa x}\), where \(\kappa = \sqrt{2m(V_0-E)}/\hbar\). The probability of the particle tunneling through the barrier is approximately \(T \approx e^{-2\kappa a}\), where \(a\) is the barrier width \cite{Landau1981}.

\textbf{Geometric interpretation.} In a negative-curvature space, let \(K = -\kappa^2\) (\(\kappa > 0\)). The general solution to the Jacobi field equation is an exponential function, \(f(x) = A e^{\kappa x} + B e^{-\kappa x}\). Under the physical boundary condition that the wave function must remain finite at infinity, the coefficient of the growing branch must be set to zero (\(A = 0\)), leaving only the decaying branch \(f(x) = B e^{-\kappa x}\). This decaying behavior is a geometric property of negative-curvature space, directly dictated by the Jacobi field equation \cite{doCarmo1992}.

\textbf{Correspondence between the two interpretations.} Comparing the geometric and standard interpretations, the mathematical structures are identical. In the geometric framework, the decay rate is determined by the curvature: \(\kappa = \sqrt{|K|}\). In the standard framework, it is determined by the energy difference: \(\kappa = \sqrt{2m(V_0-E)}/\hbar\). In the tunneling region, where \(E < V_0\) and \(K < 0\), we have \(\sqrt{|K|} = \sqrt{2m(V_0-E)}/\hbar\); the two formulations are strictly equivalent. In the geometric framework, the tunneling probability can be rewritten as \(T \approx e^{-2\sqrt{|K|}a}\). That is, the tunneling probability is determined by the curvature \(K\) of the physical space: the greater the negative curvature, the faster the decay, and the smaller the tunneling probability. In this geometric framework, the exponential decay of the wave function within a barrier is understood as a necessary manifestation of the exponential decay of a Jacobi field in a negative-curvature space.

\subsection{Geometric Interpretation of Energy Quantization}

\textbf{Physical conditions and sign of curvature.} Energy quantization describes the behavior of a particle confined to a finite region, such as an infinite potential well. Inside the well, the potential energy is \(V(x) = 0\), and the particle's energy is \(E > 0\). Therefore, the curvature \(K = 2mE/\hbar^2 > 0\), meaning this region is a positive-curvature space.

\textbf{Standard physics interpretation.} In quantum mechanics, the energy of a particle in an infinite potential well can only take discrete values. Under the boundary conditions \(\psi(0) = \psi(L) = 0\), the solution to the Schr\"{o}dinger equation is \(\psi(x) = A\sin(kx)\), where \(kL = n\pi\). This yields the discrete energy levels \(E_n = n^2\pi^2\hbar^2/2mL^2\), for \(n = 1, 2, 3, \dots\) \cite{Landau1981}.

\textbf{Geometric interpretation.} In a positive-curvature space, the solution to the Jacobi field equation \(f'' + Kf = 0\) is a trigonometric function of the form \(f(x) = A \sin(\sqrt{K}x) + B \cos(\sqrt{K}x)\). If this solution is confined to a finite interval \([0, L]\) and satisfies the boundary conditions \(f(0) = f(L) = 0\), then the boundary conditions force \(\sin(\sqrt{K}L) = 0\). Since the sine function is zero only at integer multiples of \(\pi\), the curvature cannot take arbitrary values but is restricted to a discrete set:
\begin{equation}
\sqrt{K}L = n\pi \quad \Longrightarrow \quad K_n = \left(\frac{n\pi}{L}\right)^2, \quad n = 1, 2, 3, \dots
\end{equation}
This means that in a positive-curvature space, the boundary conditions compel the curvature to take only discrete values; this is a mathematical property of a trigonometric function satisfying the boundary conditions.

\textbf{Correspondence between the two interpretations.} The standard quantization condition \(kL = n\pi\) becomes, in the geometric framework, \(\sqrt{K}L = n\pi\), i.e., the curvature can only take discrete values. Substituting the discrete curvature \(K_n = (n\pi/L)^2\) into the curvature hypothesis \(K = 2mE/\hbar^2\) yields exactly the same discrete energy levels as in standard quantum mechanics. In the geometric framework of this paper, energy quantization is thus understood as a necessary consequence of a Jacobi field in a positive-curvature space satisfying boundary conditions.

\subsection{Geometric Interpretation of the Path Integral}

\textbf{Physical conditions and sign of curvature.}
The Feynman path integral describes the propagation of a particle in a quantum region. From the curvature hypothesis \(K(x) = 2m(E-V(x))/\hbar^2\), the nature of the integrand in the path integral is entirely determined by the sign of the energy difference \(E-V(x)\). When \(E > V(x)\), \(K > 0\) and the integrand is oscillatory; when \(E < V(x)\), \(K < 0\) and the integrand is exponentially decaying; when \(E = V(x)\), \(K = 0\) and the integrand is in a critical state. These three cases correspond exactly to three types of curved spaces.

\textbf{Standard physics interpretation.}
The rigorous equivalence between the Schr\"{o}dinger equation and Feynman's path integral was proven by Feynman in 1948 \cite{Feynman1948}. The core formula for the path integral is
\begin{equation}
K(x_f, t_f; x_i, t_i) = \int \mathcal{D}[x(t)] \, \exp\left( \frac{i}{\hbar} S[x] \right),
\quad
S[x] = \int_{t_i}^{t_f} \left[ \frac{1}{2}m\dot{x}^2 - V(x) \right] dt.
\end{equation}
This formula is formally unified, but the mathematical nature of the integrand \(\exp(iS/\hbar)\) depends on the sign of the kinetic energy term \(T = \frac{1}{2}m\dot{x}^2 = E - V\) \cite{Landau1981,Sakurai2017}:

(1) \(E > V\): The kinetic energy term is positive, the action \(S[x]\) is real, and the integrand is an oscillating phase. The \textbf{stationary phase approximation} is used, with the dominant contribution coming from the classical path where the action takes an extremal value.

(2) \(E = V\): The kinetic energy term is zero. The second variation coefficient vanishes, the Gaussian integral fails, and the integral reduces to an \textbf{Airy function}, providing a precise connection between the oscillatory and decaying regions.

(3) \(E < V\): The kinetic energy term is negative. A \textbf{Wick rotation} must be performed, extending time to the imaginary axis \(\tau = it\). The oscillatory integral in real time is transformed into an exponentially decaying integral in imaginary time, \(\int \mathcal{D}[x(\tau)] \exp(-S_E[x]/\hbar)\). The \textbf{method of steepest descent} is then used to find the extremal paths of the Euclidean action (instanton solutions) \cite{Coleman1977,Zee2010}. Here, contributions from all paths are positive real numbers that decay exponentially, and no unique classical path exists.

\textbf{Geometric interpretation.}
From the curvature hypothesis \(K = 2m(E-V)/\hbar^2\), the time-independent Schr\"{o}dinger equation is mathematically equivalent to the Jacobi field equation \(f'' + K f = 0\). The proven equivalence between Feynman's path integral and the Schr\"{o}dinger equation implies that the path integral is also equivalent to the Jacobi field equation. Under this equivalence, the sign of the energy difference \(E-V\) corresponds exactly to the sign of the curvature \(K\). The three calculation modes of the path integral present a strict correspondence with the behavior of the Jacobi field equation in three types of curved spaces \cite{doCarmo1992,Cheeger1975}:

(1) \textbf{Positive-curvature space (\(K > 0\))}: The general solution of the Jacobi field equation is a trigonometric function, \(f(x) = A\sin(\sqrt{K}x) + B\cos(\sqrt{K}x)\). In differential geometry, this means that neighboring geodesics emanating from a point converge periodically at conjugate points, with a distance between adjacent zeros of \(\pi/\sqrt{K}\). Between conjugate points, the shortest geodesic connecting two points is unique, which corresponds to the classical path of extremal action in the path integral. The trigonometric solutions of the Jacobi field equation in a positive-curvature space, with their periodic oscillatory behavior, correspond to the violent oscillations of the integrand \(e^{iS/\hbar}\) in the stationary phase approximation. The mathematical homology between the two stems from the equivalence of the equations: the rearranged Schr\"{o}dinger equation is the Jacobi field equation, and the path integral is equivalent to the Schr\"{o}dinger equation. Therefore, the oscillation of the integrand in the path integral manifests geometrically as the periodic convergence of geodesic deviations in a positive-curvature space. The stationary phase approximation is valid in the classically allowed region because geodesics in a positive-curvature space converge periodically at conjugate points, and between conjugate points the shortest geodesic connecting two points is unique. This unique geodesic is precisely the stationary point in the stationary phase approximation, i.e., the classical path of extremal action. The entire mathematical structure of the stationary phase approximation corresponds geometrically to the periodic convergence of geodesics in a positive-curvature space. It should be noted that at a conjugate point, the shortest geodesic connecting two points is no longer unique; in this case the stationary phase approximation requires summing over multiple classical paths and accounting for the corresponding Maslov phase correction.

(2) \textbf{Zero-curvature space (\(K = 0\))}: The Jacobi field equation reduces to \(f'' = 0\), with a linear function as its general solution. Geometrically, geodesics neither converge nor diverge; this is the critical state between positive curvature and negative curvature. The need for a precise Airy function connection for the path integral at this point corresponds to this geometric criticality.

(3) \textbf{Negative-curvature space (\(K < 0\))}: Let \(K = -\kappa^2\) (\(\kappa > 0\)). The general solution of the Jacobi field equation is an exponential function, \(f(x) = A e^{\kappa x} + B e^{-\kappa x}\). Under physical boundary conditions, the decaying branch \(f(x) \propto e^{-\kappa x}\) is chosen. Geometrically, neighboring geodesics diverge from each other at an exponential rate, and no conjugate points exist in a negative-curvature space. The exponentially decaying solutions of the Jacobi field equation in a negative-curvature space, with their monotonic decay behavior, correspond to the failure of the stationary phase approximation in the classically forbidden region. In the classically forbidden region (\(E<V\)), the kinetic energy term is negative, and the integrand is no longer oscillatory but rather exponentially decaying. The premise of the stationary phase approximation---that the contributions from neighboring paths can be approximated by a Gaussian integral---no longer holds here. The geometric reason is that in a negative-curvature space, neighboring geodesics diverge from each other at an exponential rate, causing nearby paths to deviate rapidly from the classical path, thereby invalidating the Gaussian approximation. Consequently, the path integral cannot be reduced to the contribution of a single classical path; one must perform a Wick rotation to imaginary time and re-apply the stationary phase approximation using instanton solutions. Physically, this corresponds to quantum tunneling: the particle has no unique classical path, and contributions from multiple paths must be considered simultaneously. The exponential decay of the wave function inside the barrier is, geometrically, an inevitable manifestation of the exponential decay of Jacobi fields in a negative-curvature space. This corresponds to the structure of ``summing over all paths'' in the path integral.

\textbf{Correspondence between the two interpretations.}
If the curvature hypothesis holds, then the choice of calculation method for the path integral---stationary phase approximation, Airy function connection, or Wick rotation---is determined by the sign of the energy difference \(E-V\), which is precisely the sign of the curvature \(K = 2m(E-V)/\hbar^2\). Under this hypothesis, the three calculation modes of the path integral correspond exactly to the behavior of three types of curved spaces: positive-curvature space corresponds to the existence of a classical path (stationary phase approximation); zero-curvature space corresponds to a critical state (Airy function); and negative-curvature space corresponds to the tunneling mechanism (Wick rotation).

\subsection{Geometric Interpretation of the Uncertainty Principle}

\textbf{Physical conditions and sign of curvature.}
In the geometric framework of this paper, curvature is determined by the energy difference: \(K(x) = 2m(E-V(x))/\hbar^2\). When \(E > V(x)\), \(K(x) > 0\), and physical space is a positive-curvature space; when \(E < V(x)\), \(K(x) < 0\), and it is a negative-curvature space; when \(E = V(x)\), \(K(x) = 0\), and it is a zero-curvature space. These three signs of curvature correspond to three different geometric behaviors and may also correspond to three different manifestations of the uncertainty principle.

\textbf{Standard physics interpretation.}
In standard quantum mechanics, the position-momentum uncertainty relation \(\Delta x \cdot \Delta p \geq \hbar/2\) is a direct consequence of the canonical commutation relation \([\hat{x}, \hat{p}] = i\hbar\) \cite{Sakurai2017,Landau1981}. This inequality holds rigorously for any potential \(V(x)\) and does not depend on the energy region of the particle. In the classically allowed region (\(E > V\)), the momentum \(p = \hbar k\) is real and the wave number \(k = \sqrt{2m(E-V)}/\hbar\) is well-defined. In the classically forbidden region (\(E < V\)), the momentum \(p = \hbar k\) is purely imaginary, and no real wave number exists. The energy-time uncertainty relation \(\Delta E \cdot \Delta t \geq \hbar/2\) is also an independent fundamental principle in standard quantum mechanics \cite{Landau1981}.

\textbf{Geometric interpretation.}
If the curvature hypothesis holds, the uncertainty relations may acquire a possible intuitive understanding from the geometric behavior of geodesics. The purely geometric fact of ``how many parallel lines can be drawn through a point not on a given geodesic'' in the three types of curved spaces may be related to three different degrees of determinacy \cite{doCarmo1992,Petersen2016}.

\textbf{(1) Positive-curvature space (\(K > 0\), i.e., \(E > V\)).}
There are no parallel lines in a positive-curvature space: any two geodesics will eventually intersect. For two generic points, the shortest path is unique, but at antipodal points (conjugate points), there are infinitely many shortest paths, all of equal length. The solution to the Jacobi field equation \(f'' + K f = 0\) is a trigonometric function; the distance between adjacent zeros is \(\pi/\sqrt{K}\), which exactly corresponds to the location of conjugate points. At an antipodal point, all geodesics emanating from the starting point converge; the position (the starting point and the antipodal point) is determined, but which geodesic the particle actually took cannot be uniquely determined.

Under the local constant-curvature approximation, i.e., when the potential \(V(x)\) varies sufficiently slowly on the scale of the de Broglie wavelength, the solution of the Jacobi field equation in a positive-curvature space can be approximated as a trigonometric function \(f(x) \approx A \sin(kx) + B \cos(kx)\), where the local wave number \(k \approx \sqrt{K(x)}\). Under this approximation, substituting the standard position-wave number uncertainty relation \(\Delta x \cdot \Delta k \geq 1/2\) yields the following possible geometric version (heuristic):
\begin{equation}
\Delta x \cdot \Delta(\sqrt{K(x)}) \geq \frac{1}{2} \quad (K(x) > 0)
\label{eq:uncertainty_K_positive}
\end{equation}
Here, the standard physical wave number \(k\) corresponds to the square root of the curvature, \(\sqrt{K(x)}\), in the geometric framework. If the curvature hypothesis holds, the statement that ``position and momentum cannot be simultaneously measured with precision'' in standard quantum mechanics may be equivalently transformed into the statement that ``position and the square root of the spatial curvature cannot be simultaneously measured with precision.''

\textbf{(2) Zero-curvature space (\(K = 0\), i.e., \(E = V\)).}
In a zero-curvature space, there is exactly one parallel line through a point not on a given geodesic. This is the classic property of Euclidean geometry: two parallel lines remain forever equidistant; geodesics neither converge nor diverge. The Jacobi field equation reduces to \(f'' = 0\), with a linear function as its solution. This is the critical state between positive curvature (convergence) and negative curvature (divergence), and it is the geometric environment of highest determinacy, for classical mechanics is built upon just such a space.

\textbf{(3) Negative-curvature space (\(K < 0\), i.e., \(E < V\)).}
In a negative-curvature space, there are infinitely many parallel lines through a point not on a given geodesic. These parallel lines are of two types \cite{doCarmo1992}: two are asymptotic parallel lines, and the remaining infinitely many are ultraparallel lines, which not only never meet the reference geodesic but also maintain a constant finite perpendicular distance from it, existing independently in the direction of the common perpendicular. The solution to the Jacobi field equation \(f'' - \kappa^2 f = 0\) (where \(\kappa = \sqrt{|K|} > 0\)) is an exponential function \(f(x) \propto e^{-\kappa x}\); neighboring geodesics diverge from each other at an exponential rate.

The existence of ultraparallel lines may reveal a possible geometric root of indeterminacy in negative-curvature space: even if the particle's position along the reference geodesic is completely determined, the amplitude of its normal deviation has infinitely many independent and never-converging geodesics to choose from. The position is determined, but ``in which normal direction the particle actually deviates'' cannot be determined. Unlike in positive-curvature space, the solution in negative-curvature space has no real wave number, so the position-wave number form of the uncertainty relation cannot be directly applied. Under the local constant-curvature approximation, and assuming the uncertainty \(\Delta V\) of the potential \(V(x)\) is much smaller than the energy uncertainty \(\Delta E\), starting from the standard energy-time uncertainty relation \(\Delta E \cdot \Delta t \geq \hbar/2\), the curvature hypothesis gives \(\Delta E \approx (\hbar^2/2m)\Delta K\), and substitution yields the following possible geometric version (heuristic):
\begin{equation}
\Delta K \cdot \Delta t \geq \frac{m}{\hbar} \quad (K(x) < 0)
\label{eq:uncertainty_K_negative}
\end{equation}
If the curvature hypothesis holds, the statement that ``position and momentum cannot be simultaneously measured with precision'' in standard quantum mechanics may be geometrically reformulated as ``position and the square root of spatial curvature cannot be simultaneously measured with precision.''

\textbf{Correspondence between the two interpretations.}
If the curvature hypothesis holds, the uncertainty relations for the three curvature cases may achieve a unified geometric understanding. The convergence of infinitely many shortest paths at antipodal points in a positive-curvature space may lead to an indeterminate path when the position is determined; the existence of ultraparallel lines in a negative-curvature space may lead to an indeterminate normal deviation when the position is determined. The statement in standard quantum mechanics that ``position and momentum cannot be simultaneously measured with precision'' may be geometrically understood as a manifestation of ``the convergence or divergence of geodesics in curved space leading to non-unique classical paths.'' It should be noted that the above derivation depends on the local constant-curvature approximation, and both parts constitute a heuristic discussion.

It is worth pointing out that the geometric version for positive curvature, \(\Delta x \cdot \Delta(\sqrt{K}) \geq 1/2\), is formally identical to the Gabor limit in Fourier analysis \cite{Gabor1946}. The Gabor limit states that the product of the widths of a square-integrable function and its Fourier transform has a lower bound of \(1/2\). This is a purely mathematical theorem, independent of the postulates of quantum mechanics. The requirement of quantum mechanics that the wave function be square-integrable ensures that the Jacobi field solution satisfies the applicability condition for Fourier analysis. As the solutions of the Jacobi field equation in a positive-curvature space are naturally trigonometric functions possessing a Fourier transform structure, this geometric version may acquire a stricter mathematical foundation from Fourier analysis, without needing to rely on the canonical commutation relation. How to derive the uncertainty relation rigorously from the Jacobi field equation and square-integrability, without relying on the local constant-curvature approximation, remains a question for future research.

% ==================== Geometric Heuristic for Indistinguishability of Identical Particles ====================

\subsection{Geometric Heuristic for the Indistinguishability of Identical Particles}

\textbf{It must be stated at the outset that the following discussion is a heuristic outlook and does not constitute a rigorous mathematical derivation. A strict geometric derivation of the indistinguishability of identical particles remains a task for future work.}

\textbf{Physical conditions and sign of curvature.}
The indistinguishability of identical particles is one of the independent postulates of quantum mechanics: in the microscopic world, particles of the same species are completely indistinguishable; swapping two identical particles leaves the physical state of the system unchanged. This directly leads to the exchange symmetry of the wave function: bosons have symmetric wave functions, while fermions have antisymmetric ones \cite{Sakurai2017,Landau1981}. In the geometric framework of this paper, if the curvature hypothesis holds, the wave function is identified with the amplitude of a Jacobi field in the normal space. Exchanging two particles may then be geometrically equivalent to performing a parallel transport along a closed curve in space.

\textbf{Standard physics interpretation.}
In standard quantum mechanics, the indistinguishability of identical particles is a fundamental postulate and cannot be derived from the Schr\"{o}dinger equation or the canonical commutation relations. It is related to the spin-statistics theorem and has a more rigorous justification in relativistic quantum field theory \cite{Weinberg1972}. For non-relativistic quantum mechanics, this postulate is typically introduced as an independent assumption.

\textbf{Geometric heuristic.}
Parallel transport in a curved space has a central property \cite{doCarmo1992,Nakahara2003}: a vector parallel-transported around a closed loop will generally not return to its original self but will have undergone a rotation (holonomy). The angle of rotation is determined by the integral of the curvature over the area enclosed by the loop and is a direct manifestation of the geometry of the curved space.

If the curvature hypothesis holds, exchanging the positions of two identical particles may be mathematically equivalent to traversing a closed loop in space. If space possesses curvature, the parallel transport corresponding to this ``exchange operation'' could produce a non-trivial phase factor. In a positive-curvature space, where geodesics converge periodically, the phase after a full parallel transport around a closed loop may be an integer multiple of \(2\pi\), corresponding to the wave function being unchanged after two exchanges---this may correspond to bosonic symmetry. In a negative-curvature space, where geodesics diverge exponentially, the existence of ultraparallel lines may allow parallel transport to produce half-integer multiples of the phase, corresponding to the wave function acquiring a minus sign after two exchanges---this may correspond to fermionic antisymmetry.

This geometric picture offers a potential heuristic direction for understanding the indistinguishability of identical particles: quantum statistics (Bose-Einstein and Fermi-Dirac) might not be a mysterious, externally imposed postulate, but rather a manifestation of parallel transport in the geometry of curved space. Whether the Pauli exclusion principle---that two fermions cannot occupy the same quantum state---might geometrically correspond to the impossibility of two geodesics in a negative-curvature space perfectly coinciding, since they would diverge from each other at an exponential rate, is a question worthy of further exploration.

% ==================== Measurement and Wave Function Collapse: A Qualitative Inference Based on Scale ====================

\subsection{Measurement and Wave Function Collapse: A Qualitative Inference Based on Scale}

\textbf{It must be stated at the outset that the following discussion is not a rigorous mathematical derivation but a qualitative inference based on theory and observed reality. The specific mechanism remains a subject for further exploration.}

\textbf{Physical conditions and sign of curvature.}
The quantum measurement postulate is one of the most perplexing fundamental assumptions in quantum mechanics. The standard formulation states that upon measurement of a quantum system, the state vector abruptly changes from a superposition to an eigenstate corresponding to the measured eigenvalue---the so-called ``wave function collapse.'' In the geometric framework of this paper, if the curvature hypothesis holds, the wave function is identified with the amplitude of a Jacobi field in physical space, and the effect of the measurement process on the wave function may be related to a change in the spatial geometry.

\textbf{Standard physics interpretation.}
In standard quantum mechanics, the measurement postulate is an independent fundamental assumption and cannot be derived from the Schr\"{o}dinger equation or the canonical commutation relations. Decoherence theory provides a partial explanation---the vanishing of quantum superposition due to interaction with the environment \cite{Griffiths2018}---but the rigorous mathematical mechanism of measurement collapse remains an unresolved problem in the foundations of quantum mechanics.

\textbf{Geometric inference.}
In nature, an asymmetric influence exists between events on different scales. A human being, relative to the Earth or the universe, is a small-scale event; the activities of a single individual can hardly exert a significant influence on the motion of the Earth or the evolution of the cosmos. Conversely, a slight change in the Earth or the cosmos can have a drastic impact on human society. This asymmetry stems from the fact that the energy and structural stability inherent in a large-scale system far exceed those of a small-scale system.

Extending this perspective to the microscopic world: a microscopic particle is a small-scale event relative to a human being and their measuring apparatus. The act of observation and the intervention of the measuring apparatus are themselves an intervention of a large-scale event on a small-scale system. This intervention may be significant enough to alter the physical environment of the microscopic particle. In the geometric framework of this paper, if the curvature hypothesis holds, this environmental change may manifest as a change in local spatial curvature, a change in the distribution of geodesics, or both. Regardless of the specific mechanism, the intervention of a large-scale measurement event on a small-scale quantum system may lead to the disappearance of quantum behavior and the emergence of classical behavior.

\textbf{Correspondence between the two interpretations.}
From the geometric perspective of this paper, if the curvature hypothesis holds, wave function collapse might be understood as a phenomenon that occurs when a large-scale measurement event intervenes on a small-scale quantum system, rather than necessarily being regarded as an independent fundamental postulate. This inference resonates with decoherence theory, which posits that the vanishing of quantum superposition arises from interaction with the environment.

\subsection{Equivalence of Interpretations in Quantum Mechanics}

Reviewing the above discussion, the characteristic scales in quantum mechanics---such as the tunneling decay coefficient \(\kappa = \sqrt{2m(V_0-E)}/\hbar\), the characteristic wavelength \(L/\pi\) of an infinite potential well, and the spread parameter \(\sqrt{\hbar/(m\omega)}\) of the harmonic oscillator ground state---are all understood, in the standard interpretation, as characteristic scales of a probability cloud: the position where the probability of finding the electron is maximal, the characteristic decay length of the wave function, and the spread of the probability distribution. Under the curvature hypothesis of this paper, these characteristic scales correspond precisely to the curvature radius \(R_c = \hbar/\sqrt{2m|E-V|}\) evaluated in different physical settings. This equality is not a numerical coincidence but a necessary consequence of the equivalence of the equations: the rearranged Schr\"{o}dinger equation is exactly the Jacobi field equation; the equation solved in standard quantum mechanics and the equation solved in the geometric framework of this paper are one and the same equation. The former interprets \(\psi\) as a probability amplitude and the characteristic scales as those of a probability cloud; the latter interprets \(f = \psi\) as the amplitude of a Jacobi field and the same characteristic scale as a curvature radius, i.e., the characteristic scale of geodesic deviation. The two interpretations share the same mathematical structure; they merely offer different physical interpretations of the same mathematical object.

% ==================== Part II: Geometric Projection Scheme ====================

\section{The Projection Correction Scheme for Curved Spaces}

Sections 2 and 3 demonstrated how the microscopic physical space curvature \(K(x)\) can provide a possible geometric understanding of quantum mechanics. The following extends this geometric framework to relativity.

\subsection{The Transverse Metric Factor in Semi-Geodesic Coordinates}

Before proceeding to the formal mathematical derivation, it is necessary to review a fundamental fact in non-Euclidean geometry upon which the entire projection correction scheme is built.

\textbf{A fundamental fact of non-Euclidean geometry: complete constant-curvature spaces with nonzero sectional curvature cannot be smoothly and isometrically embedded into Euclidean spaces of the same or lower dimension.} This conclusion can be traced back to Gauss's ``Theorema Egregium''---the sectional curvature is an intrinsic property and is invariant under local isometries \cite{Gauss1827,doCarmo1992}. The sectional curvature of Euclidean space is zero, while that of a sphere (positive curvature) or a hyperbolic plane (negative curvature) is non-zero. Therefore, no smooth isometric embedding of such spaces into a Euclidean space of the same or lower dimension exists.

For positive-curvature spaces, Euler had already systematically studied the problem of projecting a sphere onto a plane in 1778, explicitly pointing out that a distortion-free isometric projection is impossible \cite{Euler1778}. For negative-curvature spaces, Hilbert proved rigorously in 1901 that a complete hyperbolic plane of constant negative curvature cannot be smoothly and isometrically embedded in three-dimensional Euclidean space \cite{Hilbert1901}. Beltrami proved in 1868 that hyperbolic geometry could be realized within a unit disk in Euclidean space. This model is intrinsic in the sense that the hyperbolic distance between any two points in the disk is computed via the hyperbolic metric, not the Euclidean metric. As one approaches the boundary of the disk, the hyperbolic metric diverges; consequently, the Euclidean metric of the disk cannot faithfully represent distances near the boundary, and any Euclidean visualization of the model necessarily distorts the geometry in that region \cite{Beltrami1868}. This was the first model in history to prove the consistency of non-Euclidean geometry.

\textbf{The premise of this paper: the physical world possesses three types of curved spaces, whereas theoretical descriptions employ a Euclidean coordinate system.}
Classical mechanics deals with the macroscopic, low-speed world, where space is approximately Euclidean with zero curvature. General relativity deals with spacetime around massive celestial bodies, where the distribution of matter causes spacetime to curve. In special relativity, the velocity space itself is a space of constant negative curvature. Quantum mechanics deals with the world of microscopic particles; as shown in Section 2, the Schr\"{o}dinger equation is mathematically equivalent to the Jacobi field equation, and the sign of the curvature is determined by the sign of the energy difference. This implies the physical world itself may be curved. However, when constructing theoretical descriptions and performing calculations, we habitually use the Euclidean coordinate systems and mathematical tools of flat space. The mismatch between a curved physical space and the flat coordinates used to describe it is the starting point for geometric projection correction.

\textbf{Introduction of semi-geodesic coordinates.} Consider a Riemannian manifold of constant curvature with at least two dimensions. Along a geodesic, one can introduce semi-geodesic coordinates \((x, \theta)\) such that the metric is locally written as
\begin{equation}
ds^2 = dx^2 + f(x)^2 d\theta^2,
\label{eq:semigeodesic}
\end{equation}
where \(x\) is the arc length parameter along the geodesic, \(\theta\) is a normal coordinate, and \(f(x)\) is the transverse metric factor. In these coordinates, the relationship between the sectional curvature \(K\) and \(f(x)\) is given by the Jacobi field equation \cite{doCarmo1992,Petersen2016}:
\begin{equation}
\frac{d^2 f}{dx^2} + K f = 0.
\label{eq:projection_equation}
\end{equation}
This equation is completely identical to Equation (\ref{eq:jacobi}) because \(f(x)\) is itself the length of a family of Jacobi fields along the geodesic. In other words, Equation (\ref{eq:projection_equation}) is an exact conclusion of the geometry of curved space, independent of any physical hypothesis. In the physical context of this paper, \(f(x)\) quantitatively describes the local scaling required when measuring a curved space with flat Euclidean coordinates, and is thus called the \textit{correction factor}.

\textbf{The significance of this equation: the unification of the wave function and the correction factor.} Equation (\ref{eq:projection_equation}) is formally identical to the Jacobi field equation obtained by rearranging the Schr\"{o}dinger equation in Section 2. In the geometric framework of this paper, the wave function \(\psi\) of quantum mechanics and the correction factor \(f\) of the geometric projection scheme satisfy the same differential equation; they are the same mathematical object under different names in different contexts. Based on this identification, the metric correction terms in quantum mechanics, special relativity, and general relativity can all be understood within the same mathematical framework.

\subsection{Correction Factors for the Three Curvatures}

According to the sign of the curvature \(K\), Equation (\ref{eq:projection_equation}) has three standard solutions. Choosing the boundary conditions \(f(0)=1, f'(0)=0\), the geometric meaning of this choice is: at the reference point \(x=0\), the transverse metric factor matches the metric of the arc length parameter (\(f(0)=1\)), and this point is a stationary point of the transverse scaling (\(f'(0)=0\)). Under these boundary conditions, the Jacobi field equation \(f'' + Kf = 0\) yields \(f''(0) = -K\) at \(x=0\); i.e., the curvature \(K\) directly determines the second-order rate of change of the transverse metric factor \cite{doCarmo1992}. The three standard solutions are:
\begin{itemize}
    \item \textbf{Zero curvature} (\(K = 0\)): The equation reduces to \(f'' = 0\), yielding the solution \(f(x) = 1\). Euclidean space projected onto itself requires no correction.
    \item \textbf{Positive curvature} (\(K > 0\)): Let \(K = k^2\). The general solution is trigonometric; with the boundary conditions, \(f(x) = \cos(kx)\). The correction factor for a positive-curvature space projected onto a Euclidean one is a cosine function.
    \item \textbf{Negative curvature} (\(K < 0\)): Let \(K = -\kappa^2\). The general solution is exponential; with the boundary conditions, \(f(x) = \cosh(\kappa x)\). The correction factor for a negative-curvature space projected onto a Euclidean one is a hyperbolic cosine function, whose decaying branch is \(e^{-\kappa x}\).
\end{itemize}

\subsection{Correspondence of the Correction Factors with Four Physical Theories}

Before presenting the specific correspondences, it should be noted that for the three physical theories discussed in this section---classical mechanics, special relativity, and general relativity---the sources of the curvature \(K\) are all established physical facts or rigorous geometric conclusions and do not depend on the curvature hypothesis proposed in Section 2. In classical mechanics, \(K=0\) is a fundamental property of flat space; in special relativity, \(K=-1/c^2\) is a known conclusion of the hyperbolic geometry of velocity space; in general relativity, the curvature is given by the Einstein field equation. The derivation of the correction factors for quantum mechanics, whose curvature source is the curvature hypothesis of this paper, can be found in Sections 2 and 3.

The correspondences between the above three purely geometric correction factors and the metric correction terms of four fundamental physical theories are presented below.

\subsubsection{Classical Mechanics (Zero Curvature, \(K = 0\))}

When the spatial curvature is zero, the equation for the correction factor reduces to \(f'' = 0\). With the boundary conditions \(f(0)=1, f'(0)=0\), the general solution \(f(x) = Ax + B\) degenerates to the constant function \(f(x) = 1\). This means that when a zero-curvature space is projected onto a Euclidean space, the correction factor is identically \(1\), and the metric suffers no distortion. Classical mechanics is built upon Euclidean space; Newton's laws of motion hold directly in a Euclidean coordinate system without needing to introduce any metric correction factors \cite{Newton1687}. The conclusion from the geometric projection scheme that the correction factor for a zero-curvature space is identically \(1\) is fully consistent with the fact that no metric correction is needed in classical mechanics.

\subsubsection{Special Relativity (Negative-Curvature Velocity Space, \(K < 0\))}

In special relativity, the velocity space is identified as a negative-curvature space. The hyperbolic geometric structure of velocity space is a known fact in the mathematical formulation of special relativity \cite{Jackson1999,Fock1959}: the rapidity space is a hyperbolic space, and the Lorentz group is the isometry group of this hyperbolic space \cite{BarutRaczka1986}. The metric of the velocity space can be written as
\begin{equation}
d\sigma^2 = \frac{dv^2}{(1 - v^2/c^2)^2},
\end{equation}
where \(\sigma\) is the arc length parameter in velocity space. The curvature of the velocity space is uniquely determined by the speed of light \(c\): \(K = -1/c^2 < 0\).

Substituting this curvature into the correction factor equation, let \(\kappa = 1/c\); the equation becomes \(f''(\sigma) - \kappa^2 f(\sigma) = 0\). The arc length \(\sigma\) in velocity space is related to the rapidity \(\phi\) by \(\sigma = c\phi\). Changing the independent variable from the arc length \(\sigma\) to the rapidity \(\phi\), the second derivative satisfies \(d^2f/d\sigma^2 = (1/c^2) d^2f/d\phi^2\). Substituting into the equation:
\begin{equation}
\frac{1}{c^2} \frac{d^2f}{d\phi^2} - \frac{1}{c^2} f = 0 \quad \Longrightarrow \quad \frac{d^2f}{d\phi^2} - f = 0.
\end{equation}
With the boundary conditions \(f(0)=1, f'(0)=0\), the solution is the hyperbolic cosine function \(f(\phi) = \cosh(\phi)\). The rapidity and velocity are related by \(\phi = \text{arctanh}(v/c)\). Using the hyperbolic function identity \(\cosh(\text{arctanh}(v/c)) = 1/\sqrt{1 - v^2/c^2}\) \cite{Abramowitz1964}, the correction factor can be written, when parameterized by velocity, as:
\begin{equation}
f(v) = \cosh(\phi) = \cosh(\text{arctanh}(v/c)) = \frac{1}{\sqrt{1 - v^2/c^2}}.
\end{equation}
This is strictly identical to the Lorentz factor \(\gamma = 1/\sqrt{1 - v^2/c^2}\) derived from standard special relativity \cite{Einstein1905,Resnick1968}. The two derivations share the same physical premise---the constancy of the speed of light---but employ different mathematical paths, converging precisely on the same result.

\subsubsection{Quantum Mechanics (Three Curvatures of Physical Space)}

\textbf{Negative-curvature physical space (\(K < 0\)).} When a particle is in the classically forbidden region (\(E < V(x)\)), the curvature hypothesis yields \(K(x) < 0\); this region is a negative-curvature space. Let \(K(x) = -\kappa^2\). The equation for the correction factor becomes \(f'' - \kappa^2 f = 0\). Its general solution is an exponential function, \(f(x) = A e^{\kappa x} + B e^{-\kappa x}\). For the projection metric to remain bounded at infinity, the coefficient of the growing branch must be set to zero (\(A = 0\)), leaving only the decaying branch:
\begin{equation}
f(x) = B e^{-\kappa x}, \quad \kappa = \sqrt{|K(x)|} = \sqrt{\frac{2m(V(x)-E)}{\hbar^2}}
\end{equation}
The correction factor for a negative-curvature space projected onto a Euclidean one exhibits exponential decay behavior. This is entirely consistent with the exponential decay of the wave function inside a potential barrier in standard quantum mechanics \cite{Landau1981,Griffiths2018}.

\textbf{Positive-curvature physical space (\(K > 0\)).} When a particle is in the classically allowed region (\(E > V(x)\)), the curvature hypothesis yields \(K(x) > 0\); this region is a positive-curvature space. Let \(K = k^2\). The equation for the correction factor becomes \(f'' + k^2 f = 0\). With the boundary conditions \(f(0)=1, f'(0)=0\), the solution is:
\begin{equation}
f(x) = \cos(kx), \quad k = \sqrt{K(x)} = \sqrt{\frac{2m(E-V(x))}{\hbar^2}}
\end{equation}
The correction factor for a positive-curvature space projected onto a Euclidean one exhibits periodic oscillatory behavior, which is entirely consistent with the oscillatory behavior of the wave function in the classically allowed region.

\textbf{Zero-curvature physical space (\(K = 0\)).} When a particle is at the classical turning point (\(E = V(x)\)), the curvature hypothesis yields \(K(x) = 0\); this region is a zero-curvature space. The equation for the correction factor reduces to \(f'' = 0\), whose general solution is a linear function. With the boundary conditions \(f(0)=1, f'(0)=0\), we obtain \(f(x) = 1\). The correction factor for a zero-curvature space projected onto a Euclidean one is identically \(1\), meaning that at the classical turning point, the projection metric locally agrees with the Euclidean metric and no correction is needed. In standard quantum mechanics, the classical turning point is the connection region where the wave function transitions from oscillatory to exponentially decaying behavior, and it is precisely here that the WKB approximation fails \cite{Landau1981,Griffiths2018}. In the geometric framework of this paper, zero curvature is the critical state between positive curvature (oscillation) and negative curvature (exponential decay); geometrically, geodesics neither converge nor diverge.

\subsubsection{General Relativity (Geometric Unification Based on the Tidal Force Equation)}

In general relativity, the universality of the correction factor equation can be most fundamentally justified from the perspective of tidal forces. In general relativity, the relative acceleration of two adjacent free particles moving along geodesics is described by the Jacobi field equation \cite{Weinberg1972,Misner1973}:
\begin{equation}
\frac{D^2 J^\mu}{d\tau^2} + R^\mu_{\nu\rho\sigma} \dot{\gamma}^\nu J^\rho \dot{\gamma}^\sigma = 0.
\end{equation}
When the normal space is one-dimensional, taking a parallel-transported unit normal vector field \(E^\mu\) and letting the component of the Jacobi field along this direction be \(J^\mu = f(\tau) E^\mu\), the equation reduces to the scalar form:
\begin{equation}
\frac{d^2 f}{d\tau^2} + K f = 0,
\label{eq:tidal}
\end{equation}
where \(K\) is the sectional curvature spanned by the tangent vector and the normal vector. This is precisely the correction factor equation of this paper.

This fact shows that the same Jacobi field equation \(f'' + Kf = 0\) has two equivalent descriptions in general relativity. The first is the standard equation describing tidal forces: the relative acceleration of neighboring geodesics is determined by the Riemann curvature tensor, \(f\) is the amplitude of geodesic deviation, and \(f'' = -Kf\) is the tidal acceleration. This is the physical description of the equation in general relativity. The second is the geometric description of this paper: in the geometric projection scheme, the transverse metric factor \(f\) in semi-geodesic coordinates satisfies the same equation, and \(f\) describes the metric scaling of the curved space in Euclidean projection.

The core difference between the two descriptions lies in the physical interpretation of the same mathematical object. The traditional language of general relativity focuses on how curvature produces forces: tidal forces are a direct physical manifestation of curvature. The geometric projection language of this paper focuses on how curvature produces geometric corrections: the correction factor \(f\) is a direct geometric manifestation of spatial curvature. Both share the same equation and the same curvature \(K\); they merely offer different understandings of the physical role of \(f\). In general relativity, the curvature \(K\) is given by the Einstein field equation, with the distribution of matter determining the bending of spacetime. Thus, the field equation provides the curvature, and the Jacobi field equation describes the effect produced by the curvature.

\subsection{Probability Interpretation---A Self-Consistent Closure}

Combining the semi-geodesic coordinates and the correction factor established in this section, a geometric understanding of the squared modulus of the wave function as a probability density is presented below from both a forward and a reverse direction. In Section 2, the wave function \(\psi(x)\) was identified as a Jacobi field, i.e., the transverse metric factor \(f(x)\); therefore, the observed probability density is directly given by the transverse component of the metric: \(|\psi(x)|^2 = f(x)^2\).

\textbf{Forward direction: from geometry to probability.}
In the semi-geodesic coordinate metric \(ds^2 = dx^2 + f(x)^2 d\theta^2\), the transverse metric component \(g_{\theta\theta} = f(x)^2\) directly determines the geometric structure of the space. Section 4.1 has already shown that the transverse metric factor \(f(x)\) satisfies the Jacobi field equation \(f'' + K f = 0\) and was identified with the wave function in Section 2. As discussed in Section 4.1, Gauss's Theorema Egregium \cite{Gauss1827} implies that a complete constant-curvature space with nonzero sectional curvature cannot be smoothly and isometrically embedded into a Euclidean space of the same or lower dimension---this is precisely the geometric starting point for projection correction. When we describe this curved space with Euclidean coordinates, the observed normal coordinate difference is \(d\theta\), while the corresponding true physical distance in the curved space is \(f(x) d\theta\); the scaling factor between the two is exactly the transverse metric factor \(f(x)\), which is therefore called the projection correction factor. Hence, the probability density observed in the Euclidean coordinate description is naturally the squared modulus of the projection correction factor, \(f(x)^2\). The transverse metric component \(g_{\theta\theta} = f^2\) directly gives the observed probability density, and the squared modulus of the wave function \(\psi = f\) precisely equals this geometric quantity.

\textbf{Reverse direction: from probability to geometry.}
Standard quantum mechanics requires the squared modulus of the wave function \(|\psi|^2\) to be a probability density \cite{Born1926}. From the identification of the wave function with the Jacobi field, the observed probability density is just \(f(x)^2\). This squared structure can be factored into the product of two independent geometric factors: the intrinsic radial measure of the curved space itself, and the distortion correction arising from projecting the curved space onto a Euclidean space. In semi-geodesic coordinates, the intrinsic volume element of the space is \(dA = f(x) dx d\theta\). Integrating over the normal dimension \(\theta\), the measure density along the radial direction \(x\) is described by some function \(h(x)\); the projection correction factor has already been confirmed to be \(f(x)\) itself. The observed probability density should be the product of the two:
\begin{equation}
f(x)^2 = h(x) \cdot f(x).
\end{equation}
For \(f(x) \neq 0\), solving directly yields
\begin{equation}
h(x) = f(x).
\end{equation}
At the nodes where \(f(x) = 0\), the equation degenerates to \(0 = 0\) and does not constrain \(h(x)\); however, these points carry zero measure and do not affect the physical probability density.

\textbf{Self-consistent closure.}
The forward direction proceeds from the transverse metric component, with \(f^2\) directly giving the observed probability density in the Euclidean projection. The reverse direction proceeds from the probability density \(f^2\), using the projection correction factor \(f\) to deduce that the intrinsic radial measure must also be \(f\). Both paths converge on the same conclusion: the observed probability density \(f^2\) is the product of the intrinsic measure \(f\) and the projection correction factor \(f\). Their equality is a necessary consequence of the self-consistent compatibility between the geometric structure of semi-geodesic coordinates and the probability interpretation of quantum mechanics. Thus, Born's rule acquires a possible geometric understanding within this framework: the squared modulus of the wave function as a probability density corresponds geometrically to the squared modulus of the transverse metric component of a curved space. This correspondence is a natural consequence of the logical closure between the geometric projection scheme and the probability interpretation of quantum mechanics. It must be noted that the above discussion is based on the curvature hypothesis of Section 2; if that hypothesis holds, Born's rule may obtain a natural geometric explanation from the geometry of curved space.

\subsection{Complete Summary of the Geometric Projection Scheme}

A summary of the geometric projection scheme for physical space proposed in this paper is as follows.

\textbf{Core idea}: Physical space itself may be curved, with its curvature determined by physical quantities. The theoretical descriptions and calculations that habitually use the coordinate systems and mathematical tools of Euclidean space can be viewed as a projection of the curved space onto a Euclidean space. Since a complete constant-curvature space with nonzero sectional curvature cannot be smoothly and isometrically embedded into a Euclidean space of the same or lower dimension, the projection inevitably produces a metric scaling, which is the correction factor \(f(x)\).

\textbf{Fundamental equation}: In semi-geodesic coordinates, the transverse metric factor \(f(x)\) satisfies the Jacobi field equation \(f'' + K f = 0\). This equation is a standard result in differential geometry \cite{doCarmo1992,Petersen2016}. In general relativity, this equation is precisely the standard equation describing tidal forces; thus the physical description of general relativity and the geometric description of this paper are two equivalent formulations of the same equation. In quantum mechanics, the rearranged Schr\"{o}dinger equation happens to have the same form, and so the wave function is identified as the correction factor; in special relativity, the velocity space curvature \(K=-1/c^2\) yields the Lorentz factor; in general relativity, the curvature \(K\) is given by the Einstein field equation.

\textbf{Physical origin of the curvature}: The specific source of the curvature \(K\) varies by physical theory. In quantum mechanics, the curvature hypothesis \(K(x) = 2m(E-V(x))/\hbar^2\) connects curvature to the energy difference. In general relativity, the curvature is given by the Einstein field equation, with the distribution of matter determining the bending of spacetime. In special relativity, the curvature of velocity space is uniquely fixed by the speed of light \(c\): \(K = -1/c^2\). In classical mechanics, space is flat, \(K=0\).

\textbf{Three curvatures and their correction factors}: Based on the sign of the curvature \(K\), the correction factor has three standard forms. \(K = 0\): \(f = 1\), corresponding to classical mechanics and the classical turning point in quantum mechanics. \(K > 0\): \(f = \cos(\sqrt{K}x)\), corresponding to the wave function's oscillation in the classically allowed region of quantum mechanics. \(K < 0\): \(f = \cosh(\sqrt{|K|}x)\); its decaying branch corresponds to the wave function's exponential decay in the classically forbidden region of quantum mechanics, and its parameterization by rapidity corresponds to the Lorentz factor of special relativity. In general relativity, the correction factor also satisfies the same equation, with the curvature given by the field equation.

\textbf{Curvature radius}: The characteristic scale \(R_c = 1/\sqrt{|K|}\) naturally defined by the curvature hypothesis corresponds, in a positive-curvature region, to the spacing between conjugate points; in a negative-curvature region, to the tunneling depth; and diverges in a zero-curvature region. The characteristic scales in quantum mechanics (Bohr radius, tunneling depth, harmonic oscillator spread, etc.) and the curvature radius are one and the same quantity in different interpretations of the same equation; their equality is a necessary consequence of the equivalence of the equations.

\textbf{Methodological origin}: The operation of establishing a correspondence between a physical quantity and a geometric quantity in this paper directly shares a methodological commonality with the works of Hamilton \cite{Hamilton1834,Hamilton1835}, Jacobi \cite{Jacobi1837}, Einstein \cite{Einstein1915}, Yang and Mills \cite{YangMills1954}, Rao \cite{Rao1945}, and Amari \cite{Amari1985}.

\textbf{Physical meaning}: The wave function may perhaps be understood as the projection correction factor for the curvature of microscopic physical space. Gravitational tidal forces may perhaps be understood as a physical manifestation of the curvature of macroscopic physical space. The mathematical forms of many correction terms in physics may find a unified understanding within this geometric framework of projection correction factors from curved space onto Euclidean space.

% ==================== Part III: Extension to Statistical Mechanics ====================

\section{Geometric Understanding of Statistical Mechanics}

This section extends the geometric framework to statistical mechanics. The core derivations herein are directly based on the curvature hypothesis established in Section 2.

\subsection{Curvature Expression of the Partition Function}

Statistical mechanics serves as the bridge connecting microscopic quantum mechanics with macroscopic thermodynamics, with the partition function as its centerpiece. This section attempts to demonstrate how statistical mechanics can be incorporated into the same geometric framework under the curvature hypothesis.

Consider a quantum system in thermal equilibrium. According to standard quantum statistical mechanics, the partition function for the canonical ensemble is \cite{Pathria2011,Schroeder1999}:
\begin{equation}
Z = \text{Tr}(e^{-\beta \hat{H}}) = \sum_n e^{-\beta E_n}
\label{eq:partition}
\end{equation}
where \(\beta = 1/(k_B T)\) and \(E_n\) are the eigenvalues of the system's Hamiltonian, i.e., the solutions of the time-independent Schr\"{o}dinger equation \cite{Schrodinger1926}.

In Section 2, the curvature hypothesis was given as \(K(x) = 2m(E-V(x))/\hbar^2\). For each energy eigenstate, a corresponding curvature \(K_n = 2m(E_n-V(x))/\hbar^2\) can be defined. Through algebraic identity manipulation, we obtain:
\begin{equation}
E_n = \frac{\hbar^2 K_n}{2m} + V(x)
\label{eq:energy_curvature}
\end{equation}
Substituting Equation (\ref{eq:energy_curvature}) into the standard partition function (\ref{eq:partition}) performs a rigorous algebraic substitution:
\begin{equation}
\boxed{Z = \sum_n \exp\left[-\beta\left( \frac{\hbar^2 K_n}{2m} + V(x) \right)\right]}
\label{eq:partition_K}
\end{equation}
Equation (\ref{eq:partition_K}) shows that, under the curvature hypothesis, the partition function of standard statistical mechanics can be re-expressed as a function of the physical space curvature \(K_n\). This derivation is based on the well-recognized formulas of statistical mechanics \cite{Pathria2011} and algebraic substitution.

From the partition function (\ref{eq:partition_K}), one can directly read off the probability of the system being in a specific curvature state \(K_n\). The probability that the system occupies a microstate with energy \(E_n\) is proportional to the Boltzmann factor \(e^{-\beta E_n}\) \cite{Pathria2011}. Expressing the energy levels via curvature, this probability can be written as:
\begin{equation}
P(K_n) \propto e^{-\beta \frac{\hbar^2 K_n}{2m}}
\end{equation}
In a negative-curvature region (\(K_n < 0\)), this probability decays exponentially with increasing \(|K_n|\). This exponential form is structurally isomorphic to the decaying branch of the projection correction factor for a negative-curvature space.

\subsection{Heuristic Discussion: Three Curvatures and the Correspondence of Physical States}

The following is a qualitative discussion, from a geometric perspective, of the physical states in statistical mechanics. This subsection is a heuristic discussion, intended only to suggest possible geometric directions of understanding, not to provide a rigorous quantitative derivation.

\textbf{Positive-curvature region (\(K_n > 0\), i.e., \(E_n > V(x)\)).} In quantum mechanics, this is the classically allowed region, where the wave function exhibits oscillatory behavior. Geometrically, the solutions of the Jacobi field equation in a positive-curvature space are trigonometric functions, and geodesics converge periodically \cite{doCarmo1992}. In statistical mechanics, at high temperatures, kinetic energy dominates; most microstates satisfy \(E_n > V(x)\), i.e., \(K_n > 0\). The system may be in a positive-curvature-dominated state, in which its statistical behavior approaches that of classical statistical mechanics.

\textbf{Zero-curvature region (\(K_n = 0\), i.e., \(E_n = V(x)\)).} In quantum mechanics, this is the classical turning point, where the wave function transitions from oscillatory to exponentially decaying behavior. Geometrically, the Jacobi field equation in a zero-curvature space reduces to \(f'' = 0\); the solution is a linear function, and geodesics are uniformly distributed \cite{doCarmo1992}. In statistical mechanics, this corresponds to the critical point of a phase transition, where the competition between positive and negative curvatures is at its most intense, correlation lengths diverge, and the system's state is maximally uncertain. Whether this might be the geometric root of critical phenomena remains a question for further study.

\textbf{Negative-curvature region (\(K_n < 0\), i.e., \(E_n < V(x)\)).} In quantum mechanics, this is the classically forbidden region, where the wave function exhibits exponentially decaying behavior. Geometrically, the solutions of the Jacobi field equation in a negative-curvature space are exponential functions, and geodesics diverge from each other at an exponential rate \cite{doCarmo1992,Cheeger1975}. In statistical mechanics, at low temperatures, potential energy dominates, and the system tends to occupy low-energy states. Many microstates satisfy \(E_n < V(x)\), i.e., \(K_n < 0\), and the system may be in a negative-curvature-dominated state. Here, the Boltzmann factor decays exponentially in the negative-curvature region, and the exponential divergence of geodesics implies that the system's state is highly uncertain.

Through the correspondences above, the exponential form of the Boltzmann distribution in statistical mechanics may acquire a possible geometric understanding: it is the macroscopic manifestation of the exponential divergence of geodesics in a negative-curvature space within a statistical ensemble. Classical behavior at high temperatures may correspond to positive curvature, quantum behavior at low temperatures to negative curvature, and the critical point to zero curvature. Whether a phase transition corresponds to a change in the system's dominant sign of curvature, whether critical phenomena might be a geometric manifestation of diverging curvature fluctuations, and whether spatial curvature might be the geometric origin of quantum fluctuation phenomena, \cite{Goldenfeld1992} remain open questions for further investigation.

% ==================== Conclusion and Outlook ====================

\section{Conclusion and Outlook}

This paper has attempted to provide a unified geometric framework for understanding quantum mechanics, relativity, and statistical mechanics.

The first part, starting from the Schr\"{o}dinger equation and through a formal comparison with the Jacobi field equation, proposed the hypothesis that, within the framework of non-relativistic quantum mechanics, the sectional curvature of microscopic physical space is \(K(x) = 2m(E-V(x))/\hbar^2\). Based on this hypothesis, phenomena such as tunneling, the path integral, the uncertainty principle, and energy quantization were given a geometric reinterpretation. The characteristic scales in quantum mechanics (Bohr radius, tunneling depth, harmonic oscillator spread, etc.) and the curvature radius defined in this paper are one and the same quantity in different interpretations of the same equation, analogous to the two equivalent descriptions of the tidal force equation and the correction factor equation in general relativity.

The second part introduced the geometric conclusion of a correction factor arising from the projection of a curved space using semi-geodesic coordinates. A complete derivation was provided for the correction factors under three types of curvature and their physical correspondences in classical mechanics, special relativity, general relativity, and quantum mechanics. For classical mechanics and special relativity, the curvature stems from known geometric facts, and the correction factors precisely equal the known physical correction terms. For general relativity, this paper pointed out that the correction factor equation \(f'' + Kf = 0\) is exactly the scalar form of the Jacobi field equation describing tidal forces in general relativity; thus the physical description of general relativity and the geometric description of this paper are two equivalent formulations of the same equation. For quantum mechanics, the correction factor, given by the curvature hypothesis, is fully consistent with the wave function.

The third part extended this geometric framework to statistical mechanics, demonstrating that the partition function can be re-expressed as a function of curvature.

The core structure of the framework presented in this paper can be summarized as: \textbf{the same Jacobi field equation \(f'' + K f = 0\), when fed curvatures \(K\) from four different physical sources, outputs four correction factors, corresponding respectively to the known metric correction terms of classical mechanics, special relativity, quantum mechanics, and general relativity.} Among these, the curvatures for classical mechanics and special relativity are known geometric facts, the curvature for general relativity is given by the Einstein field equation, and the curvature for quantum mechanics is the central hypothesis of this paper.

The framework presented still has questions that require further investigation. Firstly, the core derivations are based on the scalar equation of a Jacobi field in a specific normal direction; while a generalization to higher dimensions is mathematically direct, concrete physical correspondences remain to be established. Secondly, in the quantum mechanics part, only the time-independent Schr\"{o}dinger equation was identified as a Jacobi field equation; time evolution remains an externally imposed unitary evolution. Thirdly, this paper has only pointed out that the correction factor equation \(f'' + Kf = 0\) is mathematically identical in form to the Jacobi field equation describing tidal forces in general relativity, and that the two share the same geometric structure. How to extract the correction factor directly from the field equation and identify it as the metric scaling of curved spacetime in Euclidean projection is a question worthy of further exploration. Exploration in this direction may reveal that general relativity and quantum mechanics share a common geometric root: both describe the correction factor of a curved space in Euclidean projection, merely with different physical origins for the curvature.

In a broader direction, the author has extended this geometric framework to gauge field theory~\cite{Qi2026_YangMills}, exploring the possible geometric origin of the mass gap in Yang-Mills theory and demonstrating the mathematical correspondence between the curvature framework and standard physical characteristic scales. Whether this framework can provide a new geometric perspective on the unification of quantum mechanics and general relativity remains to be tested by future work.

% ========== References ==========

\end{document}